\documentstyle[prl,aps,floats]{revtex}
\begin{document}
\def\NOBSERVED{41}
\def\NBKG{24.8}
\def\DBKG{2.4}
\def\EXCESS{16.2}
\def\XSEC{7.1}
\def\DXSEC{2.8}
\def\DXSECSYS{1.5}
\def\TLUM{110.3}
\def\psig{0.0006}
\def\sign{3.2}
\def\cl{28.7}
\def\DOMASS{172.1}
\def\DDOMASS{7.5}
\def\NNCUT{0.85}
\def\D0XSEC{5.6}
\def\DD0XSEC{1.8}
%
\def\etal{{\sl et al.}}                 
\def\vs{{\sl vs.}}                      
\input psfig
\input epsf
\def\Aplan{$\cal{A}$}
\def\Spher{$\cal{S}$}
\def\C{$\cal {C}$}
\def\met{\mbox{${\hbox{$E$\kern-0.6em\lower-.1ex\hbox{/}}}_T$}} 
\def\ttbar{$t \bar{t}$}
\def\D0{D\O}
%
\title {Measurement of the Top Quark Pair Production Cross Section in the  
All-Jets Decay Channel}
\author{                                                                      
B.~Abbott,$^{40}$                                                             
M.~Abolins,$^{37}$                                                            
V.~Abramov,$^{15}$                                                            
B.S.~Acharya,$^{8}$                                                           
I.~Adam,$^{39}$                                                               
D.L.~Adams,$^{49}$                                                            
M.~Adams,$^{24}$                                                              
S.~Ahn,$^{23}$                                                                
G.A.~Alves,$^{2}$                                                             
N.~Amos,$^{36}$
% ...
\\                                                                            
\vskip 0.70cm                                                                 
\centerline{(D\O\ Collaboration)}                                             
\vskip 0.70cm                                                                 
}                                                                             
\address{                                                                     
\centerline{$^{1}$Universidad de Buenos Aires, Buenos Aires, Argentina}       
\centerline{$^{2}$LAFEX, Centro Brasileiro de Pesquisas F{\'\i}sicas,         
                  Rio de Janeiro, Brazil}                                     
\centerline{$^{3}$Universidade do Estado do Rio de Janeiro,                   
                  Rio de Janeiro, Brazil}                                     
\centerline{$^{4}$Universidad de los Andes, Bogot\'{a}, Colombia}             
\centerline{$^{5}$DAPNIA/Service de Physique des Particules, CEA, Saclay,     
                  France}                                                     
\centerline{$^{6}$Panjab University, Chandigarh, India}                       
\centerline{$^{7}$Delhi University, Delhi, India}                             
\centerline{$^{8}$Tata Institute of Fundamental Research, Mumbai, India}      
\centerline{$^{9}$Kyungsung University, Pusan, Korea}                         
\centerline{$^{10}$Seoul National University, Seoul, Korea}                   
\centerline{$^{11}$CINVESTAV, Mexico City, Mexico}                            
\centerline{$^{12}$Institute of Nuclear Physics, Krak\'ow, Poland}            
\centerline{$^{13}$Institute for Theoretical and Experimental Physics,        
                   Moscow, Russia}                                            
\centerline{$^{14}$Moscow State University, Moscow, Russia}                   
\centerline{$^{15}$Institute for High Energy Physics, Protvino, Russia}       
\centerline{$^{16}$University of Arizona, Tucson, Arizona 85721}              
\centerline{$^{17}$Lawrence Berkeley National Laboratory and University of    
                   California, Berkeley, California 94720}                    
\centerline{$^{18}$University of California, Davis, California 95616}         
\centerline{$^{19}$University of California, Irvine, California 92697}        
\centerline{$^{20}$University of California, Riverside, California 92521}     
\centerline{$^{21}$Florida State University, Tallahassee, Florida 32306}      
\centerline{$^{22}$University of Hawaii, Honolulu, Hawaii 96822}              
\centerline{$^{23}$Fermi National Accelerator Laboratory, Batavia,            
                   Illinois 60510}                                            
\centerline{$^{24}$University of Illinois at Chicago, Chicago,                
                   Illinois 60607}                                            
\centerline{$^{25}$Northern Illinois University, DeKalb, Illinois 60115}      
\centerline{$^{26}$Northwestern University, Evanston, Illinois 60208}         
\centerline{$^{27}$Indiana University, Bloomington, Indiana 47405}            
\centerline{$^{28}$University of Notre Dame, Notre Dame, Indiana 46556}       
\centerline{$^{29}$Purdue University, West Lafayette, Indiana 47907}          
\centerline{$^{30}$Iowa State University, Ames, Iowa 50011}                   
\centerline{$^{31}$University of Kansas, Lawrence, Kansas 66045}              
\centerline{$^{32}$Louisiana Tech University, Ruston, Louisiana 71272}        
\centerline{$^{33}$University of Maryland, College Park, Maryland 20742}      
\centerline{$^{34}$Boston University, Boston, Massachusetts 02215}            
\centerline{$^{35}$Northeastern University, Boston, Massachusetts 02115}      
\centerline{$^{36}$University of Michigan, Ann Arbor, Michigan 48109}         
\centerline{$^{37}$Michigan State University, East Lansing, Michigan 48824}   
\centerline{$^{38}$University of Nebraska, Lincoln, Nebraska 68588}           
\centerline{$^{39}$Columbia University, New York, New York 10027}             
\centerline{$^{40}$New York University, New York, New York 10003}             
\centerline{$^{41}$University of Rochester, Rochester, New York 14627}        
\centerline{$^{42}$State University of New York, Stony Brook,                 
                   New York 11794}                                            
\centerline{$^{43}$Brookhaven National Laboratory, Upton, New York 11973}     
\centerline{$^{44}$Langston University, Langston, Oklahoma 73050}             
\centerline{$^{45}$University of Oklahoma, Norman, Oklahoma 73019}            
\centerline{$^{46}$Brown University, Providence, Rhode Island 02912}          
\centerline{$^{47}$University of Texas, Arlington, Texas 76019}               
\centerline{$^{48}$Texas A\&M University, College Station, Texas 77843}       
\centerline{$^{49}$Rice University, Houston, Texas 77005}                     
}                                                                             

\maketitle
%
\begin{abstract}
We present a measurement of $t\bar{t}$ production in $p\bar{p}$ collisions at 
$\sqrt{s}=1.8$ TeV from 110 pb$^{-1}$ of data collected
in the all-jets decay channel with the \D0~detector at Fermilab.
A neural network analysis yields a cross section of 7.1 $\pm$ 2.8 (stat.) 
$\pm$ 1.5 (syst.)~pb, at a top quark mass ($m_t$) of \DOMASS~GeV/$c^2$. 
Using previous D\O~measurements from dilepton and single lepton channels, 
the combined D\O\ result for the \ttbar~production cross section is 
5.9~$\pm$ 1.2~(stat.) $\pm$ 1.1 (syst.)~pb for $m_t$~=~\DOMASS~GeV/$c^2$.
\end{abstract}
%
\twocolumn
%
\newpage
	
The standard model predicts that, at Tevatron energies, top quarks are 
produced primarily in $t\bar{t}$ pairs, and that each top quark decays into 
a $b$ quark and a $W$ boson. 44\% of these events are expected to have both
$W$ bosons decay into quarks. 
These pure hadronic, or ``all-jets'', $t\bar{t}$ events are among the
rare collider events with several quarks in the final state. 
With no final state energetic neutrinos, the
all-jets mode is the most kinematically constrained of the top quark decay
channels, but is also the most challenging to measure due to the large QCD
multijet background.
This compelled us to use unique tools such as quark/gluon jet differences,
and to make extensive use of neural networks, to separate the $t\bar{t}$ 
final states from the QCD background\cite{normetal}. 
The comparison of $t\bar{t}$ cross sections from the all-jets and 
lepton + jets channels allows a search for new phenomena in top decays; 
for example, top decay via a charged Higgs boson could be observed as a 
deficit, relative to the all-jets final states, in the $t\bar{t}$ final 
states with energetic leptons. 

The signal for these all-jets $t\bar{t}$ events is at
least six reconstructed jets. The main background is from 
QCD multijet events that arise from a 2$\rightarrow$2 parton process
producing two energetic (``hard'') leading jets and less energetic (``soft'') 
radiated gluon jets. 
	

The \D0~detector is described in Ref. \cite{d0detector}. We used
the same reconstruction algorithms for jets, muons, and electrons as
those used in previous top quark analyses\cite{ouroldprd}. The muons
in this analysis are used to identify $b$ jets, and are restricted to the 
pseudorapidity range $|\eta|\leq~1.0$, where 
$\eta = {\rm tanh}^{-1}({\rm cos}\theta)$,
and $\theta$ is the polar angle relative to the beam axis. 


The multijet data sample was selected using a hardware trigger 
and an online filter requiring five
jets of cone size $\cal{R}$ = 0.5, pseudorapidity $|\eta|< 2.5$ and 
transverse energy $E_T > 10.0$ GeV. Here, $\cal{R}$ = $((\Delta\phi)^2
+ (\Delta\eta)^2)^{\frac{1}{2}}$, where $\phi$ is the azimuthal angle around 
the beam axis. Additionally, we required the total transverse energy of the 
event ($H_T$) to be $>$ 115 or 120 GeV (depending on run conditions). The 
data sample after the initial cuts has $\approx$ 600,000 
events. With about 200 expected top events in this channel, the background 
overwhelms the signal by a factor of $\approx$ 3000. As discrimination from 
many variables was needed to separate signal from background, most of which
are significantly correlated, we used neural networks (NN) as an integral 
part of this analysis. 


The offline analysis proceeded by excluding events with an isolated muon or 
electron to maintain a data sample independent of the other $t\bar{t}$ 
samples. We required events to have at least 
six $\cal{R}$=0.3 cone jets and less than nine $\cal{R}$=0.5 cone jets,
with jet $E_T > 8.0$ GeV.
We generally used $\cal{R}$=0.3 cone jets
because of their greater reconstruction efficiency,
but used $\cal{R}$=0.5 cone jets to calculate mass-related variables. 
We required that at least one jet have an associated muon which satisfied
muon quality criteria and which was kinematically consistent with a 
$b \rightarrow \mu X$ decay within the jet. As about 20\% 
of $t\bar{t}$ all-jets events have such a ``$\mu$-tagged'' jet in 
the acceptance region for $t\bar{t}$ signal, compared to approximately 3\% 
of the QCD multijet background in that region, the tagging requirement 
reduces the background-to-signal ratio by about an order of magnitude. 
Of the total 280,000 events surviving the offline cuts, 3853 have at least one 
$\mu$-tagged jet. These tagged events comprise the data sample used to 
extract the cross section.

Compared with the QCD multijet background,
$t\bar{t}$ events typically have more energetic jets, 
have the total energy more uniformly distributed among the jets, 
are more isotropic, and have their jets distributed at smaller $\eta$. To
discriminate $t\bar{t}$ signal from QCD background, we defined at
least two variables describing each of these qualities 
(total energy, jet energy distribution, event shape, and rapidity 
distribution)\cite{normetal}:
  
\begin{enumerate}
\setcounter{enumi}{0}

\item $H_T$: The sum of the transverse energies of jets.\\
\vspace{-0.2in}

\item $\sqrt{\hat{s}}$: The invariant mass of the jets in the final state.\\
\vspace{-0.2in}

\item $E_{T_1}$/$H_T$: $E_{T_1}$ is the transverse energy of the leading jet.\\
\vspace{-0.2in}

\item $H_T^{3j}$: $H_T$ without the transverse energy of the two leading 
jets.\\
\vspace{-0.2in}

\item $N_{\rm jets}^A$: The number of jets averaged over a range of $E_T$ 
thresholds (15 to 55 GeV), and weighted by the $E_T$ threshold.
This parameterizes the number of jets taking their hardness into account.\\
\vspace{-0.2in}

\item $E_{T_{5,6}}$: The square root of the product of the transverse energies 
of the fifth and sixth jets.\\
\vspace{-0.2in}

\item ${\cal{A}}$: The aplanarity, calculated from the normalized momentum
tensor.\\
\vspace{-0.2in}

\item ${\cal{S}}$: The sphericity, calculated from the normalized momentum
tensor.\\
\vspace{-0.2in}

\item ${\cal{C}}$: The centrality, ${\cal{C}} = 
{H_T}/{H_E}$, where $H_E$ is the sum of all the jet total energies. This 
characterizes the transverse energy flow.\\
\vspace{-0.2in}

\item $<$$\eta^2$$>$: The $E_T$-weighted mean square of the $\eta$ 
distribution of jets in an event.\\
\vspace{-0.2in}
\end{enumerate}
\noindent

These ten variables are the inputs to the first neural network (NN1), 
whose output is used as an input variable for the second neural network (NN2).
The three other inputs to NN2 are:

\begin{enumerate}
\setcounter{enumi}{10}

\item $p_T^{\mu}$: The transverse momentum of the tagging muon. 
\end{enumerate}

The $p_T^{\mu}$ distribution is harder for tagged jets in 
$t\bar{t}$ events than for tagged jets in QCD multijet events. 

\begin{enumerate}
\setcounter{enumi}{11}
\item $\cal{M}$: The mass-likelihood variable. This variable is defined as
${\cal M} = (M_{W_1}-M_{W})^2/\sigma_{m_W}^2~+~(M_{W_2}-M_{W})^2/\sigma_{m_W}^2
~+~(m_{t_1}-m_{t_2})^2/\sigma_{m_t}^2$, with the parameters $M_W$, 
$\sigma_{m_W}$, and $\sigma_{m_t}$ set to 80, 16 and 62 GeV/$c^2$, respectively.
$M_{W_i}$ and $m_{t_i}$ refer to the jet combinations that best define 
the $W$ boson and top quark masses in an event. 
\end{enumerate}

The mass likelihood variable $\cal{M}$ is a $\chi^2$-like quantity, 
minimized when there are two invariant masses consistent with the $W$ mass, 
and two candidate top quark masses that are identical. $\sigma_{m_W}^2$ 
and $\sigma_{m_t}^2$ were determined from simple two and three 
jet combinations using D\O\ jet resolutions. We did not assume that the muon
tagged jet came from a $b$ quark. 

\begin{enumerate}
\setcounter{enumi}{12}

\item $\cal{F}$: The jet-width Fisher discriminant. This is defined as
${\cal F}_{\rm jet} = (\sigma_{\rm jet}-\sigma_{\rm quark}(E_T))^2/
\sigma^{2}_{\rm quark}(E_T) - (\sigma_{\rm jet}-\sigma_{\rm gluon}(E_T))^2/
\sigma^{2}_{\rm gluon}(E_T)$, where $\sigma^{2}_{\rm quark}(E_T)$ and
$\sigma^{2}_{\rm gluon}(E_T)$ are mean square jet widths calculated from 
{\sc herwig}\cite{herwig} Monte Carlo, for quarks and gluons respectively, 
as functions of jet $E_T$. 
\end{enumerate}

It has been demonstrated that quark jets are, on
average, narrower than gluon jets \cite{KEK,LEP1}. The Fisher discriminant, 
based on the $\eta$-$\phi$ {\sc rms} jet widths, is calculated for the
four narrowest jets in the event, and indicates
whether the jets were most probably ``quark-like'' ($t\bar{t}$) or 
``gluon-like'' (QCD multijet). 

\begin{figure}
\centerline{\psfig{figure=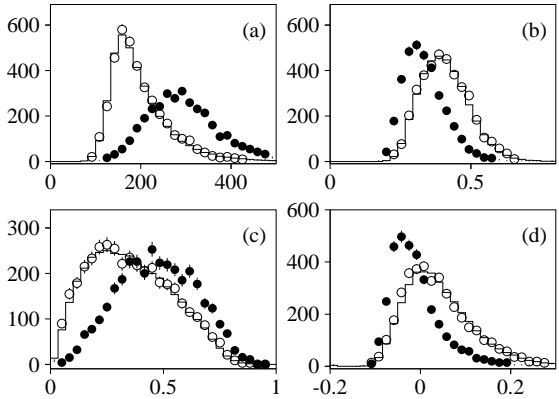,width=3in}}
\caption[Comparisons.]
{\small Comparison of the absolute number of data-based multijet background 
(histogram) with the observed 3853 muon-tagged events in data ($\circ$) for 
(a) $H_T$ (in GeV), (b) $E_{T_1}/H_T$, (c) sphericity, and (d) jet-width
Fisher discriminant. Shown also are the distributions in these variables for
{\sc herwig} $t\bar{t}$ events ($\bullet$).}
\label{muonpred}
\end{figure}

Figure~\ref{muonpred} shows a comparison of distributions from the 
modeled background discussed below, the data, and {\sc herwig} 
$t\bar{t}$ events for four of the above variables.
	
The top quark production cross section is calculated from the 
output of NN2. Both networks were trained to force their output near 
1 for $t\bar{t}$ events, and near 0 for QCD multijet events, using the 
back-propagation learning algorithm in {\sc jetnet} \cite{jetnet}. 

The very large background-to-signal ratio in the untagged data indicates an 
almost pure background sample. With a correction for the very small 
$t\bar{t}$ component expected, and with a method of assigning a 
muon tag to the untagged event, the background estimate can be determined 
directly from the data. Separate sets of untagged data with added muon tags 
were used for network background response training and background modeling.
{\sc herwig}~$t\bar{t}$ events were used for 
the $t\bar{t}$ network signal response training.

The correct assignment of muon tags to the untagged data was critical 
to our background model. We derived a ``tag rate function'' from the entire 
multijet data set, defined as the probability for any 
individual jet to have a tagging muon. We chose a function that factorized
into two pieces: $\epsilon$, the detector efficiency dependent on $\eta$ of 
the jet and the run number of the event (to account for chamber aging), and 
$f(E_T)$, the probability that a jet of tranverse energy $E_T$ has a tagging 
muon. We studied two parametrizations of $f(E_T)$, and used the 
difference to estimate the systematic error from this source. Finally, a 
small dependence of the tag rate on $\sqrt{\hat{s}}$ of the event was found, 
which was incorporated into $f(E_T)$. A detailed discussion of the 
tag rate function is given in Ref.~\cite{normetal}.

We established that the $p_T$ of the tagging muon and the $E_T$ of the 
tagged jet (uncorrected for the muon and neutrino energy) are 
uncorrelated. Therefore, the muon $p_T$ factors out of the tag rate 
function, and can be generated
independently. By applying the tag rate function to each jet  
in the untagged data sample, and generating a muon $p_T$ for those jets
determined as tagged, we produced the background model sample. 


\begin{figure}
\centerline{\psfig{figure=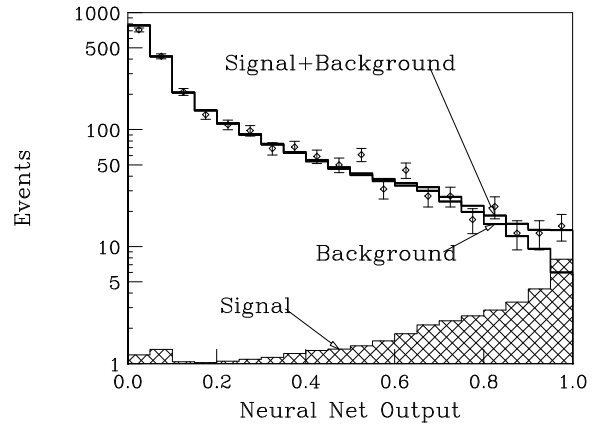,width=3in}}
\caption[Combined fits]
{\small The distribution in NN2 output (on a log scale) for data 
(diamonds + error bars) and the fits for expected signal and background.
 The signal was modeled using {\sc herwig} for $m_t$=180~GeV/$c^2$. 
The errors shown are statistical.}
\label{comb_fit}
\end{figure}

The NN2 output distributions for data, modeled background and {\sc herwig}
$t\bar{t}$ signal are plotted in Fig.~\ref{comb_fit}.
We excluded events in the region of NN2 output $<$ 0.02. Jets in that region 
tend to have low $E_T$, where the tag rate is not well determined due to the 
low tagging probability (low statistics), and consequently, the background 
modeling may be less accurate. 
The cross section is obtained from a simultaneous fit of the data to the 
background and {\sc herwig} $t\bar{t}$ shapes, with the background 
normalization ($A_{\rm bkg}$) and the $t\bar{t}$ cross section 
($\sigma_{t\bar{t}}$) as free parameters. The result of this fit is 
also shown in Fig.~\ref{comb_fit}. 

\begin{figure}
\centerline{\psfig{figure=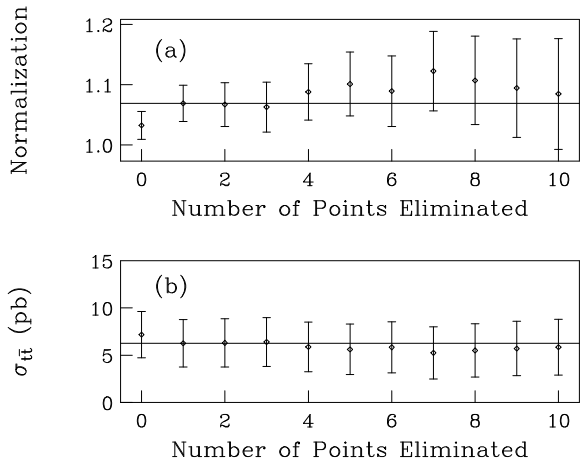,width=3in}}
\caption[Combined fit Errors]
{\small The (a) background normalization and (b) $t\bar{t}$ cross section 
from fits to the NN2 output distribution as data points in Fig.~\ref{comb_fit} 
are removed at low NN2 output values. The horizontal lines are the calculated
normalization and cross section, respectively, from Figure~\ref{comb_fit}.
Error bars are statistical, but are correlated through the error matrix.}
\label{comb_err}
\end{figure}

The stability of our result can be checked by successively
eliminating data points at the lowest values of the NN2 output.
Figure~\ref{comb_err} shows the values of the background normalization
and $t\bar{t}$ cross section as the data points are removed and the 
remaining points are refitted. The refitted cross sections are independent 
of NN2 output region, confirming that the initial NN2 output cut at 0.02, 
and choosing the region NN2 output $>$ 0.1 for our final cross section
calculation, does not bias the result. Because of the preponderance of 
background at the low end of NN2 and the stability of our fits, we use the 
region NN2 $>$ 0.1 for our quoted cross section results.

Values of the cross section and background normalization are obtained 
from similar fits with {\sc herwig} $t\bar{t}$ events generated at 
different top quark masses. The results are shown in Table I. 
Interpolating to the top quark mass as measured by D\O~\cite{snydermass} 
($m_t$ = 172.1 GeV/$c^{2}$), we obtain
$\sigma_{t\bar{t}}$ = \XSEC~$\pm$ \DXSEC~(stat.)~$\pm$ \DXSECSYS~(syst.)~pb,
consistent with a previous measurement in this channel~\cite{cdfalljets},
and the most precise value for this channel to date.
Table II summarizes the contributions to the systematic error on the
cross section. These were determined by varying each source by its 
uncertainty, and calculating the difference in the cross section.

As a check, we calculated the cross section from the excess events 
over expected background, using the efficiency of the criteria 
for $t\bar{t}$ selection (calculated using {\sc herwig}), along with the
branching ratio and the measured luminosity. For NN2 $>$
\NNCUT~(chosen to minimize the error on
the cross section) we observed \NOBSERVED~events with \NBKG~ $\pm$ \DBKG~ 
expected background events for an excess of 16.2 events. The excess 
corresponds to a $t\bar{t}$ cross section of 7.3~$\pm$~3.3~$\pm$~1.6 pb at 
$m_t$ = 172.1 GeV/$c^{2}$, consistent with our result above.

\begin{table}
 \begin{small}
 \caption{\small Results of the fits to neural network output.}
 \begin{center}
 \begin{tabular}{c|c|c|c}
 Top Quark               & $A_{\rm bkg}$   &  $\sigma_{t\bar{t}}$     
& $\chi^2$ / d.o.f.  \\
 Mass (GeV/$c^2$)        &                 &  (pb)                    
&                 \\
 \hline
 140   &  1.05 $\pm$ 0.03  &  18.4 $\pm$ 7.8  &  17.6 / 17 \\
 160   &  1.06 $\pm$ 0.03  &   9.3 $\pm$ 3.8  &  17.2 / 17 \\
 170   &  1.07 $\pm$ 0.02  &   7.2 $\pm$ 3.0  &  17.1 / 17 \\
 180   &  1.07 $\pm$ 0.03  &   6.3 $\pm$ 2.5  &  16.9 / 17 \\
 200   &  1.07 $\pm$ 0.03  &   5.1 $\pm$ 2.0  &  16.8 / 17 \\
 220   &  1.07 $\pm$ 0.03  &   4.4 $\pm$ 1.7  &  16.7 / 17 \\
 \end{tabular}
 \end{center}
 \end{small}
\end{table}

The significance of the excess is characterized by the probability $P$ of the 
observed number of events being due to fluctuation. For an NN2 output threshold
of $\approx$ 0.94, where Monte Carlo studies predict maximal expected  
significance, we observe 18 events where 6.9~$\pm$~0.9 background events 
are expected, for which $P$ = \psig, corresponding to a \sign~standard 
deviation effect. This is sufficient to establish the existence of a 
$t\bar{t}$ signal in the all-jets final state.

To further check the validity of the tag rate function and hence the
background model, we looked at events with more than one tagged jet. 
The modeled background here consists of those untagged events that
had two jets tagged by application of the tag rate function. We assumed
that the fraction of the double-tagged events from correlated sources,
such as $b\bar{b}$ production, is constant over the NN2 output, but refitted
the background normalization for a possible overall correlation.
A total of 32 double-tagged events are observed for NN2 output $>$ 0.02 where
28.7~$\pm$~8.2 events are expected from background. Two events are observed 
for NN2 output $>$ \NNCUT~with 0.7~$\pm$~0.1 expected background events, and
1.2 top events expected from Monte Carlo. 
The small excess in the double-tagged sample is consistent 
with our conclusion that the more significant
excess in the singly-tagged sample is from $t\bar{t}$ production.

Previous D\O~measurements of \ttbar~production  
in the dilepton and single lepton channels \cite{meenajim}
give an average cross section of 5.6~$\pm$ 1.4~(stat.) $\pm$ 
1.2~(syst.)~pb at $m_t$=\DOMASS~GeV/$c^2$, 
in very good agreement with that from the all-jets 
channel. We combine the all-jets cross section with these results, 
assuming the statistical errors are uncorrelated, and that the systematic 
errors have the appropriate correlation coefficients. 
The combined D\O~result for the \ttbar~production cross section is 
5.9~$\pm$ 1.2~(stat.) $\pm$ 1.1 (syst.)~pb for $m_t$=\DOMASS~GeV/$c^2$.

\begin{table}
\caption[Summary of statistical and systematic uncertainties.]
{\small Summary of statistical and systematic uncertainties for 
the cross section.}
\begin{center}
\begin{small}
\begin{tabular}{l|r}
Background Source & Size of Uncertainty   \\
\hline
Statistical error                    &  4 \% \\
Functional Form of the Muon-Tag Rate &  7 \% \\
Background Correction for $t\bar{t}$ Signal  & 6 \% \\
Background $E_T$ scale  & 9 \%   \\
\hline
Signal Source & Size of Uncertainty   \\
\hline
Statistical Error          & 3 \%  \\
Trigger Turn-on            & 5 \%  \\
Luminosity Error           & 5 \%  \\
Jet Energy Scale           & 6 \%  \\
$t\bar{t}$ Tag Rate       & 7 \%  \\
Model Dependence           & 6 \%  \\
$b\rightarrow\mu$ Branching Fraction     & 6 \%  \\
muon $p_T$ Dependence     & 7 \%  \\
${\cal{F}}$ Dependence     & 2 \%  \\
\end{tabular}
\end{small}
\end{center}
\label{systematictable2}
\end{table}

%
We thank the Fermilab and collaborating institution staffs for
contributions to this work and acknowledge support from the
Department of Energy and National Science Foundation (USA),
Commissariat  \` a L'Energie Atomique (France),
Ministry for Science and Technology and Ministry for Atomic
   Energy (Russia),
CAPES and CNPq (Brazil),
Departments of Atomic Energy and Science and Education (India),
Colciencias (Colombia),
CONACyT (Mexico),
Ministry of Education and KOSEF (Korea),
and CONICET and UBACyT (Argentina).
%

\end{document}